\documentclass{preprint}
\input{acat-hepdata-preamble.tex}

\title{HepData reloaded: reinventing the HEP data archive}

\ShortTitle{HepData reloaded}

\author{Andy Buckley\\
  School of Physics and Astronomy\\
  University of Edinburgh\\
  E-mail: \email{andy.buckley@ed.ac.uk}}

\author{Mike Whalley\\
  Institute of Particle Physics Phenomenology\\
  Durham University\\
  E-mail: \email{m.r.whalley@cern.ch}}

\abstract{%
  We describe the status of the HepData database system, following a major
  re-development in time for the advent of LHC data. The new HepData system
  benefits from use of modern database and programming language technologies, as
  well as a variety of high-quality tools for interfacing the data sources and
  their presentation, primarily via the Web. The new back-end provides much more
  flexible and semantic data representations than before, on which new external
  applications can be built to respond to the data demands of the LHC
  experimental era. The HepData re-development was largely motivated by a desire
  to have a single source of reference data for Monte Carlo validation and
  tuning tools, whose status and connection to HepData we also briefly review.
}

\FullConference{13th International Workshop on Advanced Computing and Analysis Techniques in Physics Research, ACAT\,2010\\
  February 22-27, 2010\\
  Jaipur, India}

\begin{document}

\section{Introduction}

The HepData ``reaction database'' is a repository of data from particle and
nuclear physics experiments dating back to the 1960s, which has been hosted at
Durham University since the 1970s. The emphasis of the database content is on
published distributions -- that is, it neither attempts to store raw
experimental data nor particle properties such as branching fractions, which are
the preserve of the complementary Particle Data Group~\cite{Amsler:2008zz}.
Approximately 10000 papers are archived in HepData.

Historically, HepData was originally accessed via teletype, then interactive
shell terminal. Finally, a Web interface built on CGI scripts was introduced in
the early 1990s: this is the interface to which most users are accustomed. In
this note, we describe a major upgrade effort from 2005--2009 which rebuilt the
entire HepData infrastructure and migrated all existing data with significant
improvements in functionality and data quality -- in time for the LHC
experimental era.

This upgrade was part of the CEDAR~\cite{Buckley:2007hi} e-Science project, and
was largely motivated by the desire to programmatically obtain and synchronise
reference data for HEP Monte Carlo generator tuning from HepData. The new
HepData system provides the mechanisms for this and other uses of data from our
community's experimental legacy and for the LHC era.

We now describe the implementation of the new HepData system.

\section{Object model}

Database systems for complex data structures can rapidly become unmaintainable
if all data access requires knowledge of the database table structure. Loose
coupling of data storage and usage was adopted to avoid lock-in to either a
single DB engine or a flawed initial DB schema design.  A programmatic object
model of HEP data structures was developed, with a persistency layer to
translate between the objects and their database representation. The object
model allows application code to concentrate on clearly expressing its
data-handling logic, rather than the details of how the data is retrieved or
stored.

The most important object in the HepData model is the \kbd{Paper}, representing
an experimental publication and containing a collection of \kbd{Dataset}
objects, each representing a plot or table in the publication. Each \kbd{Paper}
also references a collection of metadata summarising its publication history,
run conditions, featured reactions, principal authors, etc., which are used for
database searches, in particular via the Web interface. Each \kbd{Dataset}
contains at least one \kbd{XAxis} and at least one \kbd{YAxis} object,
representing the combination of controlled variables and sets of measurements
corresponding to those configurations respectively. Each \kbd{XAxis} contains
several \kbd{Bin} objects and each \kbd{YAxis} contains several \kbd{Point}s:
these provide access to bin ranges, widths, foci, etc., and to various data
uncertainties respectively. All numerical quantities are represented in a
semantic units system, allowing consistent automatic unit conversions in
applications while rendering the data in the units preferred by the authors of
the source papers. Consistency through the object model is ensured by reciprocal
parent--child contracts between objects, so that all object relationships can be
traversed in both directions.

Java was chosen as the implementation language for the object model due to its
object orientation and high-level standard and external libraries. The Maven 2
software control system was successfully used to handle builds of all parts of
the HepData system, including automatic management and retrieval of all
third-party libraries.

The result of this development is that database searching can now be performed
to the axis level rather than the paper, with numerical values decoupled from
their textual representation by the units system. Combined with persistency
layers to file and database storage, this structure is the core of new applications
using the HepData information.

The main remaining restriction in the object model is its adherence to a model
with a single data-type: binned distributions. Hence, it is not easy to
meaningfully represent data where the $x$-values have no obvious bin edges, such
as often is the case with collider energies, sets of data points where the
$x$-errors do not correspond to bin widths, or data where there is no natural
$x$ axis at all\footnote{A dummy $x$-axis value and error, usually the collider
  energy with errors of $\pm 1$, is currently used as a work-around in this
  case.}. There is also no natural representation for more abstract data objects
such as bin-error correlation matrices. Extension of the data types to handle
several distinct types of data object would allow for semantically correct
storage of a wider range of experimental data.

\section{Object persistency}

Hierarchical relationships such as those in HepData can be expressed in
relational form for use by SQL databases such as MySQL or Oracle, by
implementation of parent--child links -- but maintaining consistency and
navigating efficiently requires care. In HepData, this job is performed by the
Java Hibernate~\cite{hibernate} framework.  Ownership relationships between
objects are declared via an extended form of Java Persistency Annotations (JPA),
which abstract most of the details of table representation. Hibernate also
avoids object orphaning, ensures that db transactions are atomic, and helps to
optimise database querying by judicious use of lazy and eager fetching
strategies. The optimisation of these features must be done manually, but is at
a much higher level than manually implementing the database accesses. The
Envers~\cite{envers} database versioning layer may be used in conjunction with
Hibernate at some point, to allow for arbitrary roll-back of database contents.

File persistency is internally an important feature of the new HepData system,
particularly since the migration of the 10000 papers from the legacy database
was performed via intermediate files. Because of its ubiquity and natural
hierarchical structure, an internal XML dialect -- HepML -- was developed for
representation of HepData records, implemented via the external Castor
marshalling framework~\cite{castor}. This data format is not intended for
external use, since the diversity of HepData records makes the format rather
lengthy: it is instead intended for database backups, and as an intermediate
format for data input and updates.

\section{Migration}

The migration of data records from the old HepData system to the new was a
highly non-trivial task, largely due to the lack of strict structure in the
legacy database. This was centered on the hierarchical Berkeley Database System,
an unmaintained DBMS accessed programmatically from Fortran routines and without
a concept of field type: all stored values are strings. Accordingly many records
in the legacy database were found to be corrupted with typographical errors or
unparseable spacing introduced for presentational purposes: these problems were
too numerous for manual correction, and had to be resolved as part of the
migration process to the new data structures.

This migration was performed by dumping the legacy database to a set of flat
text files -- one for each data type -- via a combination of Fortran and Perl
scripts. These files were then merged into one HepML file per logical paper
using a Python script, \kbd{mkhepml}, which cross-references legacy papers to
build complete data sets while heuristically correcting for the aforementioned
misformattings, typos, and other improvable elements.

\kbd{mkhepml} makes substantial use of the BeautifulSoup~\cite{beautifulsoup}
Python XML library -- designed for tolerant screen-scraping, but used because of
its powerful and flexible element-searching API. To speed up the system, lazy
searching for object hierarchies was introduced, combined with pre-build sorting
of the flat file entries: this step minimised the number of XML tag searches
required. At this stage, most entries would take several seconds to build, but a
substantial number of large papers would take many minutes or even several hours
to migrate to HepML. This problem was obviated by introduction of a
multi-threading layer so that slow papers would not delay the entire processing
queue. In the final form, migration of the entire database could be performed
overnight: necessary to keep up with the arrival of new papers in the legacy
database, and fixes to corrupt records which were only discovered as a result of
the migration process.

This phase of HepData re-development proceeded for several years, both because
of the time taken to develop a fully-working object model and Web interface and
because of the intrinsic scale of the migration process. All new records are now
added, and updates made, directly on the new database and the migration process
from the legacy database has been happily decommissioned.

\section{Web interface}

The new HepData Web interface, like the rest of the runtime system, is written
in Java. The Tapestry 5~\cite{tapestry5} framework is used rather than raw Java
Servlets, since it combines a high-level approach to building Web applications
with strong Hibernate and Maven 2 integration, and excellent scalability.

The resulting Web interface is designed around the idea that the URL should
itself be a clean and predictable user interface: each area of the URL
corresponds to a Java page-rendering class which interprets the rest of the URL
as a set of arguments and produces a bytestream of HTML, XML, text or graphic as
appropriate: multiple output formats are trivial to add, given a routine which
constructs the desired representation from the object model -- in fact, this can
even be done by third-party applications via read-only access to the
database.

Hence, URLs under the \kbd{view} page can be programmatically restricted to
point to any (valid) combination of axes on any single dataset of a paper, and
that data can be represented as e.g. HTML tables, HepML, plain text, ROOTCINT,
or PyROOT macros. The \kbd{view} renderer also makes use of a \kbd{plot}
renderer, which dynamically produces data plots of arbitrary size in PNG or PDF
formats, with all combinations of log or linear axes -- all programmatically
accessible and bookmarkable via the predictable URL scheme. POST-type HTML form
handling is specifically avoided for this reason. Internally, a simple
templating markup language (TML) allows static HTML elements to be written as
HTML, with substitutions delegated to Java classes, allowing re-use of common
code components between pages.

\begin{figure}[tbp]
  \centering
  \includegraphics[width=0.7\textwidth]{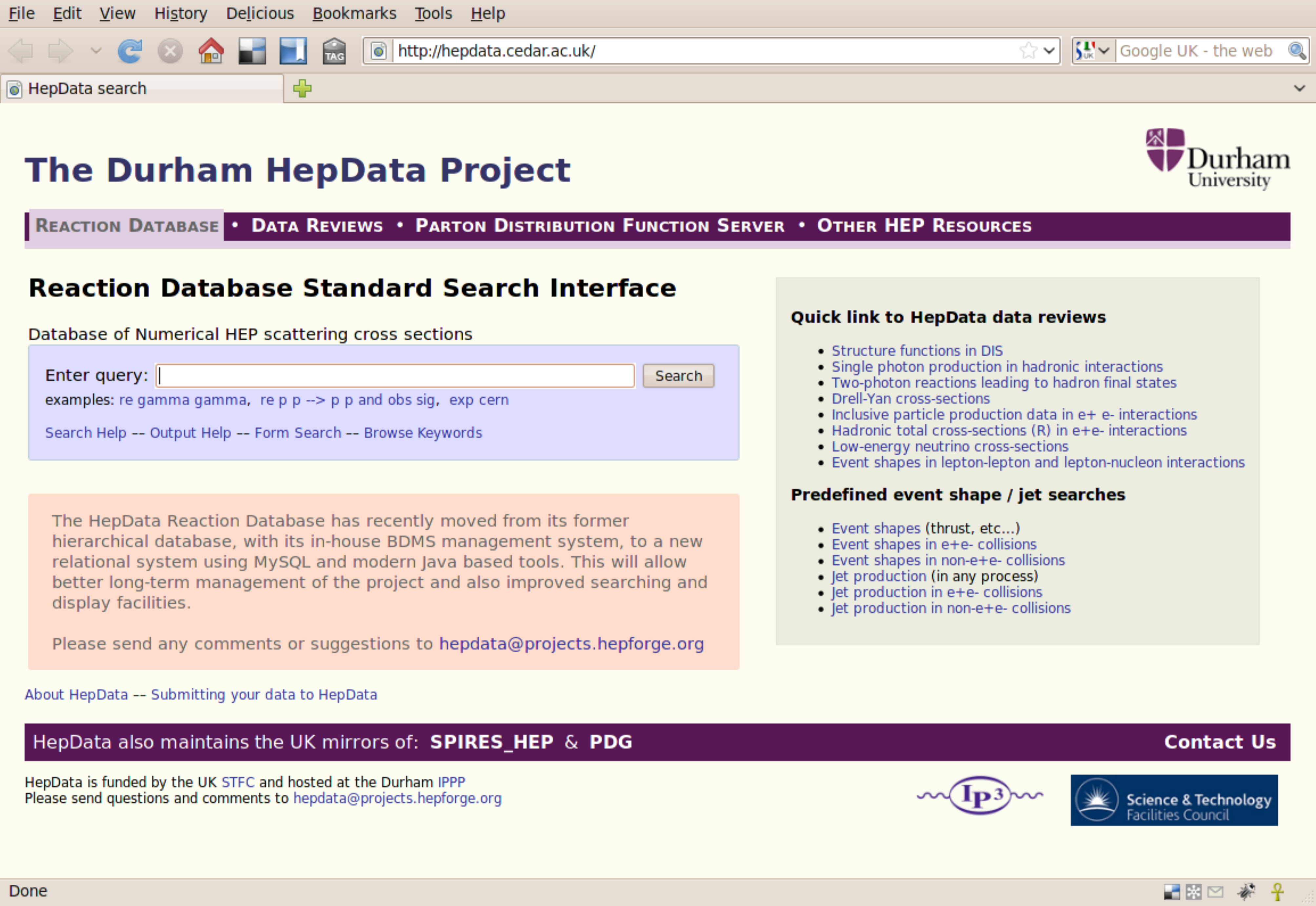}
  \caption{HepData main Web search page}
  \label{fig:hdfront}
\end{figure}

\begin{figure}[tbp]
  \centering
  \includegraphics[width=0.7\textwidth]{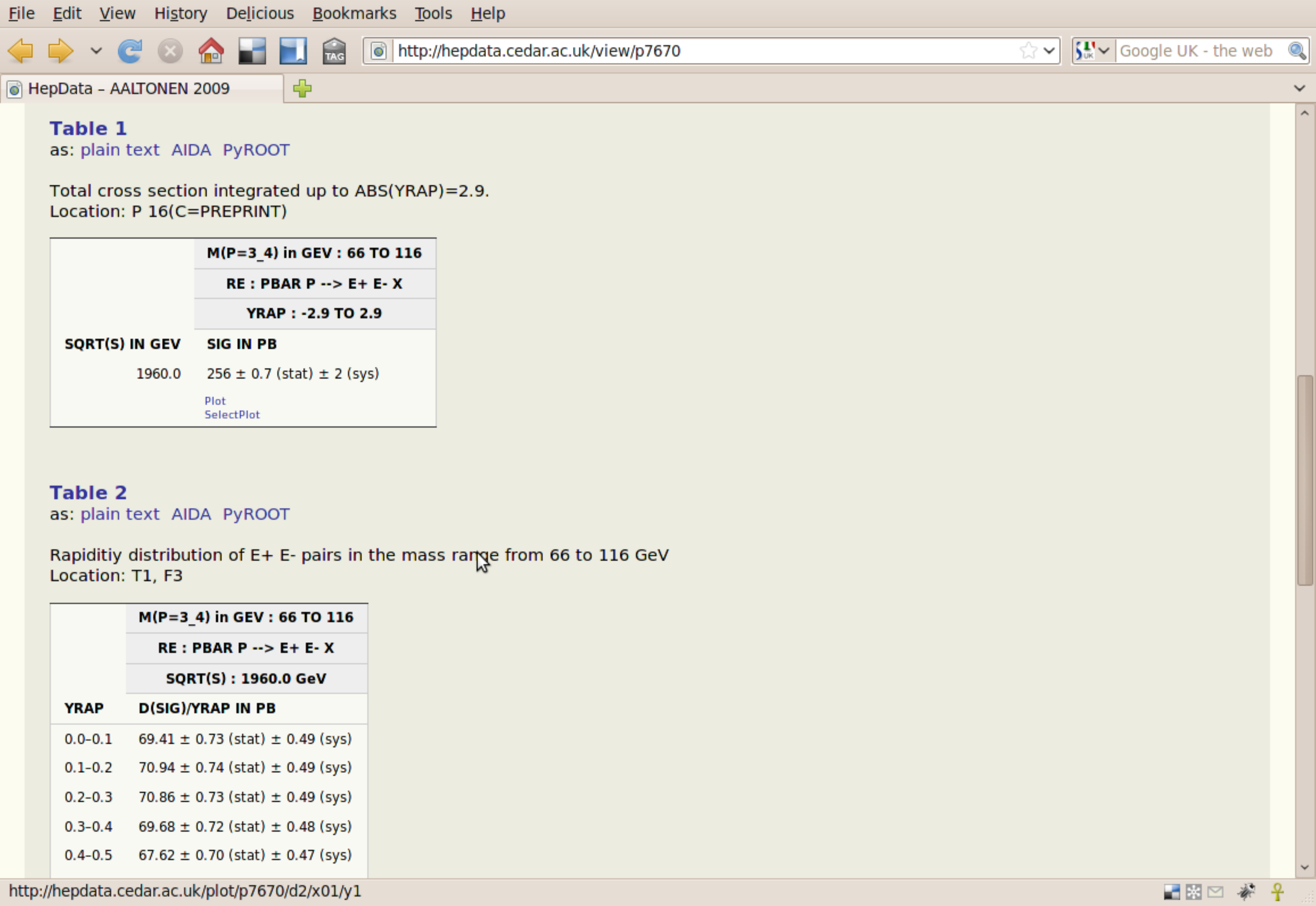}\\
  \caption{HepData paper display as HTML tables}
  \label{fig:hdtables}
\end{figure}

\begin{figure}[tbp]
  \centering
  \includegraphics[width=0.7\textwidth]{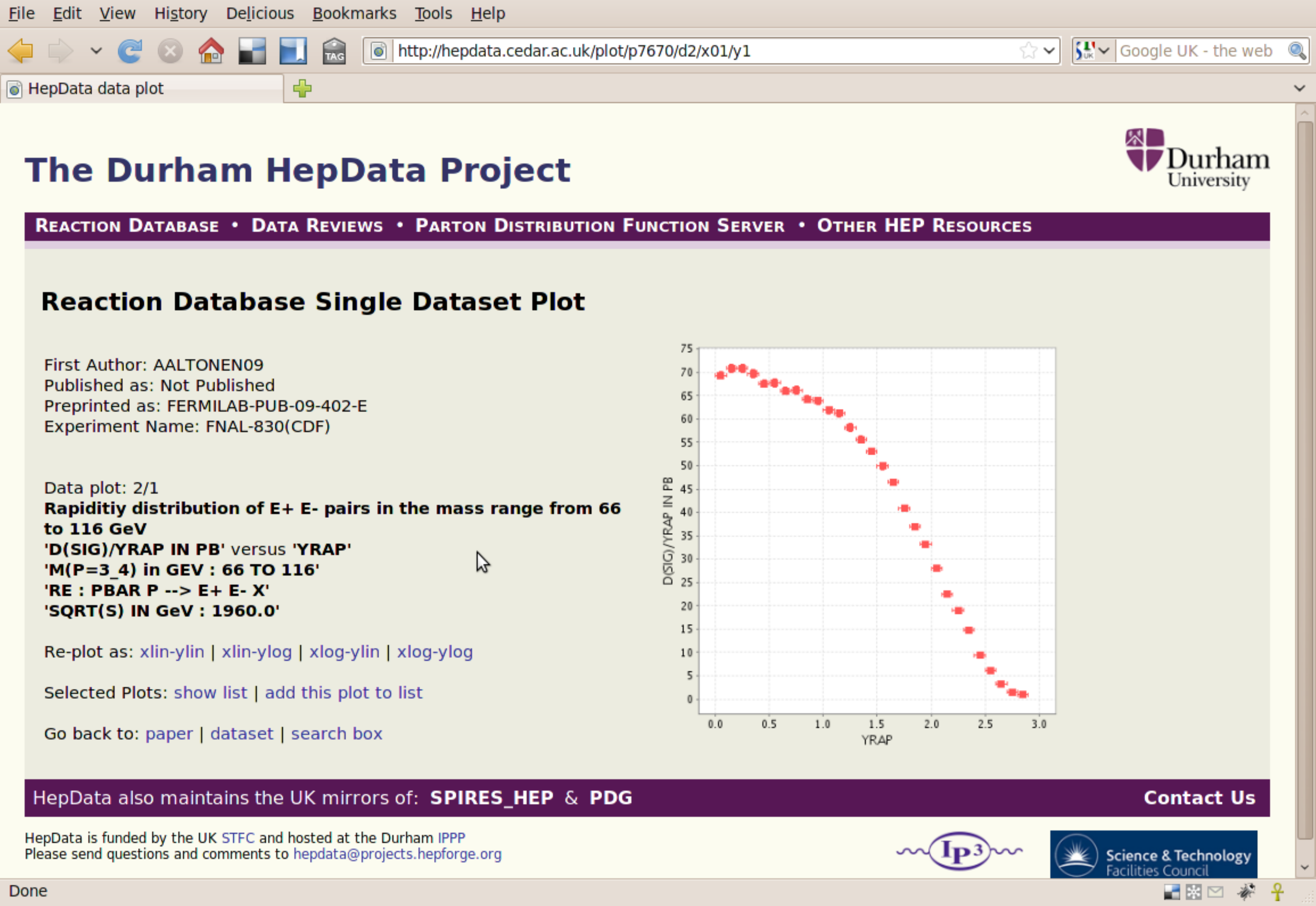}\\[0.5em]
  \caption{HepData plot display via dynamic rendering}
  \label{fig:hdplot}
\end{figure}

The HepData server system consists of a Jetty Java servlet container and MySQL
database sitting behind a firewall and exposed via a reverse proxy on an Apache
2 Web server. The system is shared with the popular HepForge~\cite{hepforge} HEP
code development system.  The HepData system is live at
\url{http://hepdata.cedar.ac.uk}, and screenshots of major interface elements
are shown in Figures~\ref{fig:hdfront}--\ref{fig:hdplot}.

The first LHC data was successfully submitted to HepData after some discussion
of submission formats: work is underway via LCG to provide tools for conversion
of ROOT data objects to a suitable (simplified) input format for HepData and for
pre-submission validation of the generated data files.

\section{HepData in MC generator validation and tuning}

The re-development of HepData was largely driven by the requirements of the MC
generator community for validation and optimisation infrastructure. Reliable
reference data is required for quality generator tuning, and a perennial problem
has been the translation errors during transcription of tabulated data into
arrays in validation code, exacerbated by the lack of any synchronisation
between the reference data hard-coded into the analyses and the paper itself,
which may receive post-publication updates. The development of a more flexible
HepData interface means that this data synchronisation can now be far more
direct, exemplified by the Rivet~\cite{Buckley:2010ar} validation system.

Rivet is a generator-agnostic analysis framework, intended as a re-engineering
of the \textsc{Fortran} HZTool library with the benefits of hindsight and a more
expressive application programming language. Rivet has two main r\^oles:
\begin{itemize}
\item to provide a collection of physics utility routines, such as jet algorithms and event shapes;
\item to collect a set of analyses which produce data comparable with experimental results.
\end{itemize}
Rivet is implemented in object oriented C++ and specifically only accesses event
data via the HepMC event record~\cite{Dobbs:2001ck} -- it is hence generator-independent by
design. Other features ensure the scalability and robustness of the system.

The reference data in Rivet are bundled with the package as AIDA XML files, one
for each paper. These files can be directly imported from HepData by an
automated URL query using the paper's SPIRES identifier as a unique key. Inside
the Rivet package, the HepData records are used to automatically choose the
binnings for histogrammed observables -- this makes the Rivet analysis code far
more compact and readable than its predecessors, and reference data fixes and
updates can be applied without code changes.

Rivet is currently being used by Genser and LHC experiments as a validation
tool, and by the Professor~\cite{Buckley:2009bj} tuning framework as the main
source of data for event generator tuning. Professor performs MC parameter
optimisations by first parameterising generator responses to parameter
variations as a large set of analytic functions, and then numerically minimising
a goodness-of-fit measure -- again, the reference data in this goodness of fit
is provided automatically by HepData. Professor is a joint project between
Edinburgh University, Durham University (IPPP), Lund University, and Berlin
Humboldt University as part of the MCnet research network, and is used by
phenomenologists and by the LHC experiments for their internal MC tuning
efforts.

\section*{Acknowledgements}
We would like to thank James Monk for presenting this material in our
absence. HepData development has been supported by a long-term grant from the UK
Science and Technology Facilities Council, and its re-development by the CEDAR
e-Science project, also supported by STFC. A.~Buckley additionally acknowledges
support from an STFC Special Project Grant, from the Scottish Universities
Physics Alliance (SUPA), and from the EU MCnet Marie Curie Research Training
Network (funded under Framework Programme 6 contract MRTN-CT-2006-035606) for
both financial support and for many useful discussions and collaborations with
its members.

\bibliographystyle{h-physrev3}
{\raggedright
  \bibliography{refs}
}

\end{document}